\documentclass[twocolumn,showpacs,preprintnumbers,amsmath,amssymb,prl]{revtex4}

\usepackage{graphicx}
\usepackage{dcolumn}
\usepackage{bm}

\begin{document}
\title{Shear-melting of a hexagonal columnar crystal by proliferation of dislocations}
\author{Laurence Ramos and Fran\c{c}ois Molino}
\affiliation{Groupe de Dynamique des Phases Condens\'{e}es (UMR
CNRS-UM2 5581), CC26, Universit\'{e} Montpellier 2, 34095
Montpellier Cedex 5, France}

\email{ramos@gdpc.univ-montp2.fr}
\date{\today}

\begin{abstract}
A hexagonal columnar crystal undergoes a shear-melting transition
above a critical shear rate or stress. We combine the analysis of
the shear-thinning regime below the melting with that of
synchrotron X-ray scattering data under shear and propose the
melting to be due to a proliferation of dislocations, whose
density is determined by both techniques to vary as a power law of
the shear rate with a $2/3$ exponent,  as expected for a creep
model of crystalline solids. Moreover, our data suggest the
existence under shear of a line hexatic phase, between the
columnar crystal and the liquid phase.
\end{abstract}

\pacs{83.60.Rs, 61.30.St, 61.10.Eq, 61.72.Hh}
\maketitle


Understanding the behavior of complex fluids under flow is
essential for their processing and technical use \cite{Larson}.
The relevance of non-linear rheology for a wide range of
applications has motivated many fundamental studies, both
theoretical and experimental. However, the rheology of structured
fluids is not well understood yet, despite robust experimental
facts such as shear-banding associated with marked modifications
of the structure or texture. This very general behavior is
observed in a large variety of materials including amorphous,
crystalline or liquid crystalline systems. Among liquid crystals,
two-dimensional ($2$D) solids, or hexagonal columnar crystals,
have been the subject of very few studies, but these materials
should deserve a particular attention because of both applicative
and fundamental grounds. Their close analogy, in terms of
elasticity and phase diagrams, to $2$D magnetic flux line lattices
(FLL) in type II high-$\rm{T_{c}}$ superconductors is appealing.
On the other hand, as columnar liquid crystals are extensively
used for the synthesis of mesoporous materials \cite{Mesoporous},
understanding and in turn controlling their structure under flow
may lead to some technological advancements.

We use synchrotron small-angle X-ray scattering (SAXS) and
rheology to investigate the behavior under shear of a soft
hexagonal columnar crystal, which consists of oil tubes arranged
on a triangular lattice in water. Rheological experiments show a
shear-induced transition between two states of markedly different
viscosities. The low viscosity high shear rate structure is a $2$D
liquid of tubes, thus indicating that, above a critical shear rate
or shear stress, a shear can induce the melting of the long-range
$2$D order of the tubes \cite{Melting}. We exploit the analogy
with FLL to derive a scenario for the melting of the soft
hexagonal columnar crystal and propose the melting to be due to a
proliferation of dislocations. We combine the analysis of the
shear-thinning regime below the melting transition and that of the
SAXS patterns under shear to extract the shear rate dependence of
the density of dislocations, $\rho$. Both techniques show that
$\rho$ increases with the shear rate as a power law with the same
exponent $2/3$, as expected for a creep model of crystalline
solids. The two independent determinations are fully consistent
and hint at a dislocation-mediated melting of the hexagonal
columnar crystal under shear, in noticeable analogy with
theoretical predictions for FLL \cite{Corbino}. Our data moreover
suggest the occurrence of an intermediate hexatic phase under
shear.

\begin{figure}
\includegraphics{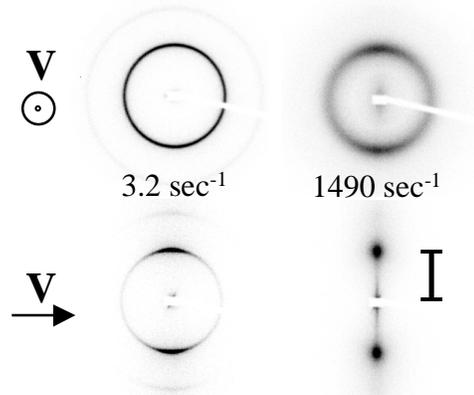}
 \caption{$2$D diffraction
patterns for a shear rate (left) $\dot{\gamma}=3.2 \,
\rm{sec}^{-1}$ and (right) $\dot{\gamma}=1490 \, \rm{sec}^{-1}$ in
(top) tangential configuration with incident beam parallel to the
velocity \textbf{V}, (bottom) radial configuration with incident
beam parallel to the velocity gradient. Scale bar is $0.2 \,
\rm{nm^{-1}}$.}.
 \label{FIG:1}
\end{figure}

\begin{figure}
\includegraphics{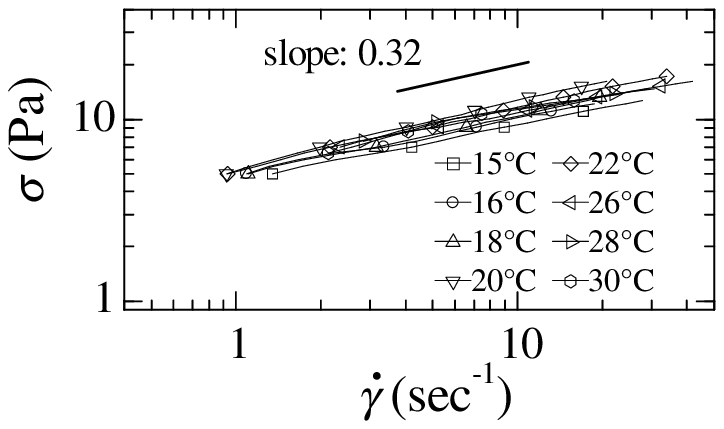}
 \caption{Flow curve, stress versus shear rate, in the low shear
rate regime. Curves are labeled by temperature.}.
 \label{FIG:2}
\end{figure}

The experimental system is a lyotropic hexagonal  phase consisting
of infinitely long oil tubes immersed in water. The samples are
composed of a quaternary mixture of sodium dodecylsulfate (SDS),
pentanol, cyclohexane and brine with a NaCl concentration of $0.4
\rm{M}$. We use a composition in weight percent of $9.7\%$ SDS,
$24.1\%$ brine, $61.5\%$ cyclohexane and $4.7\%$ pentanol which
yields, at rest, a structure consisting of oil tubes of radius $15
\, \rm{nm}$, coated with a surfactant monolayer and arranged on a
triangular array with a lattice parameter $a_{0}=33 \, \rm{nm}$
\cite{Swelling}. Rheology experiments are performed in a Couette
geometry with a stress-controlled Paar Physica UDS 200 rheometer.
The flow curve, stress $\sigma$ \textit{vs} shear rate $\dot{
\gamma}$, of a soft hexagonal columnar crystal, has been described
in detail in Refs.\cite{Melting,HexagoTcc}. At low shear rates
($\dot{\gamma}\leq\dot{\gamma}_{c}$) the system behaves as a power
law shear-thinning material while at high shear rates, the stress
varies affinely with the shear rate. These two regimes correspond
to two stable branches of stationary states, for which data
obtained by imposing either $\sigma$ or $\dot{\gamma}$ exactly
superpose. The transition from the lower branch to the higher
branch occurs through a stable hysteretic loop in a
stress-controlled experiment. To characterize the structure of the
material under flow, SAXS experiments under controlled shear rate
in Couette geometry are performed on the ID-2 beamline at the
ESRF, Grenoble, France \cite{ESRF}. The structure of the sample in
the two stable branches has been previously characterized
\cite{Melting}. Examples of the $2$D patterns obtained in both
tangential and radial configurations are given in Fig.\
\ref{FIG:1}. At low shear, the hexagonal phase exhibits a
polycrystalline texture with the tubes progressively aligned along
the velocity direction as $\dot{\gamma}$ increases, although a
state with the tubes perfectly aligned along the flow is never
reached. The angular width of the arcs (Fig.\ \ref{FIG:1}, bottom)
is used as a measure of the alignment and indicates a sample
mosaic spread, $\Delta \theta$, which decreases monotonically from
$26 \, \rm{deg}$ to $15 \, \rm{deg}$ when the shear rate increases
from $1$ to $100 \, \rm{sec^{-1}}$. By contrast, at high shear
rate, the long-range two-dimensional order of the tubes melts,
leading to a $2$D liquid of tubes strongly aligned along the flow
($\Delta \theta \simeq 8 \, \rm{deg}$). Melting transitions under
shear have been frequently observed for three dimensional
colloidal crystals \cite{shearmeltingexp} and are generally
believed to be due to the existence of a periodic potential in the
direction of flow \cite{shearmeltingtheo}. Similar arguments can
certainly not be invoked for a $2$D solid, for which a crystalline
array of tubes perfectly aligned along the velocity should flow at
low viscosity. A novel approach is thus needed.

To better understand the physical mechanism for the shear-melting,
the behavior of the material under moderate shear is analyzed
within the framework of work hardening  of crystalline solids. The
parallel between the behavior under shear of liquid crystal phases
and that of metals has been proved successful in the case of
lamellar systems \cite{Kleman}. In the formal theory of work
hardening, the stress required to move a dislocation through a
region having a density of dislocation $\rho$ can be derived from
general dimensional arguments \cite{HardeningTheo}. With the
assumption that the applied stress, $\sigma$, is relaxed by the
dislocations, $\sigma$ should depend solely on three parameters,
the density of dislocation, $\rho$, the Burger vector of the
dislocations, $b$ and the shear modulus of the material, $G$. The
pure number $\sigma /G$ must therefore be a function of the pure
number $b\rho^{1/2}$. On the other hand, the force which $\sigma$
exerts on a unit length of a dislocation, the Peach and Koehler
(PK) force, is $b\sigma$.  This force is balanced by a resisting
force arising from the line tension, which is proportional to
$b^{2}$. In a stationary state, $b \sigma \sim b^{2}$. Thus,
$\sigma$ varies linearly with $b$ and a relation between the
stress and the density of dislocation follows:

\begin{equation}
\sigma=KGb\rho^{1/2}   \label{stressvsrho}
\end{equation}

where $K$ is a numerical factor. Equation (\ref{stressvsrho}) is
well obeyed by crystalline solids, with the prefactor $K$ of the
order of $2$ for metals \cite{HardeningExp}. For hexagonal
columnar phases, there are three generic types of dislocation
\cite{deGennes}: a screw dislocation, a (longitudinal) edge
dislocation with tangent $t$ parallel to the columns dislocation
and a (transverse) edge dislocation with $t$ perpendicular to the
columns. The transverse edge dislocation is the most costly in
energy because its creation requires the formation of two end-caps
and is thus presumably very rare in soft columnar crystals. By
contrast, longitudinal edge and screw dislocations do not require
the creation of end-caps and are certainly more easily created;
these dislocations are expected to form loops with both
longitudinal edge and screw components, as also assumed in
Ref.\cite{Nelson} for FLL. For these two types of dislocation
\cite{Nelson}, the line tension is proportional to $b^{2}$. Thus,
Eq. (\ref{stressvsrho}) should hold and an increase of the density
of dislocation with the applied stress is predicted. Moreover,
whenever dislocation motion is the dominant plastic deformation
mechanism, one observes a constant shear rate regime usually
described by Orowan's geometrical relation, $\dot{\gamma}=\rho bv$
\cite{creep}, where $v$, the average velocity of the dislocations,
is proportional to the applied stress $\sigma$: $v=M\sigma$, with
$M$ a mobility. The shear rate dependence of both the stress and
the density of dislocations can thus be extracted from the
combination of Orowan's relation and Eq. (\ref{stressvsrho}). One
obtains: $\sigma=(\frac{bG^{2}}{M})^{1/3}\times
\dot{\gamma}^{1/3}$ and $\rho=(bGM)^{-2/3}\times
\dot{\gamma}^{2/3}$. We are able to measure independently the
shear rate dependences of both $\sigma$ and $\rho$. As we shall
see below, we find $\rho \sim \dot{\gamma}^{2n}$ and $\sigma \sim
\dot{\gamma}^{n}$ with  $n=1/3$, in remarkable agreement with  the
simple dimensional theory described above.

\begin{figure}
\includegraphics{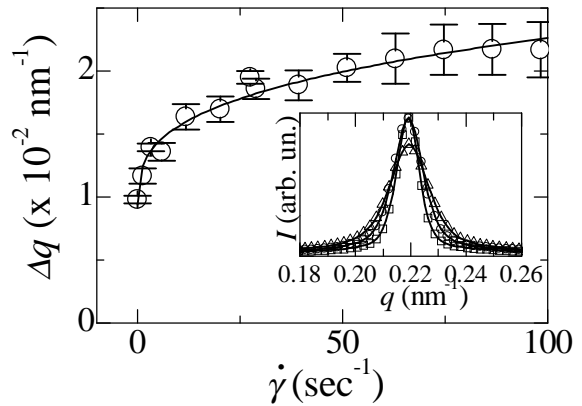}
\caption{Shear rate dependence of the full width at half height of
the first diffraction peak of the sample. The solid line is the
best fit of the data (see text). Insert: First diffraction peak
for a sample at rest (squares), submitted to a shear rate
$\dot{\gamma}=5.8 \, \rm{sec}^{-1}$ (circles) and $\dot{\gamma}=29
\, \rm{sec}^{-1}$ (triangles).} \label{FIG:3}
\end{figure}

Rheology measurements give directly the shear-thinning behavior of
the sample. As can be seen in Fig.\ \ref{FIG:2}, in a large range
of temperature, from $15$ to $30 \, ^{\circ}\rm{C}$, and over more
than one order of magnitude for $\dot{\gamma}$, we find $\sigma
\sim \dot{\gamma}^{n}$ with $n=0.32\pm0.03$, which implies $\rho
\sim \dot{\gamma}^{0.64\pm0.06}$. The shear rate variation of the
density of dislocation can be confirmed by independent
measurements. Thanks to the Synchrotron high resolution,
quantitative information on $\rho$ can indeed be obtained from the
width of the diffraction peaks. The peaks are not resolution
limited and the full width at half height (FWHH) of the first
order diffraction peak, $\Delta q$, increases with $\dot{\gamma}$,
while its position remains unchanged. Figure \ref{FIG:3} shows
that the FWHH ranges from $10^{-2}\, \rm{nm}^{-1}$ at rest up to
$2.2 \times 10^{-2}\, \rm{nm}^{-1}$ for $\dot{\gamma}=100\,
\rm{sec}^{-1}$. We note that this curve is fully reversible and in
particular the zero shear value of $\Delta q$ is systematically
recovered when the shear is stopped. We find that the variation of
$\Delta q$ can be very well fit by the sum of a constant and a
power law of the shear rate: $\Delta q=\Delta q_{0}+
g(\dot{\gamma})$, where the constant $\Delta q_{0}$ is related to
the size of the crystallites at rest. We obtain $\Delta q_{0}=0.9
\times 10^{-2}\, \rm{nm}^{-1}$ and $g(\dot{\gamma})=
B\dot{\gamma}^{m}$, with $B=0.3 \times 10^{-2}\, \rm{nm}^{-1}$ and
$m=0.33\pm 0.06$. Finite size effects can account for the shear
rate dependence broadening of the diffraction peaks. The standard
Laue-Scherrer relation relates indeed the width of the Bragg peaks
to $\xi$, a translational correlation length:
$g(\dot{\gamma})=2\pi / \xi$. In the case of a crystalline solid,
$\xi$ can be identified as the average distance between
dislocations \cite{notepolycrystal} and thus $\rho=\xi^{-2}$. From
the power law variation of $g$, one predicts the density of
dislocation $\rho$ to scale as
$\dot{\gamma}^{2m}=\dot{\gamma}^{0.66 \pm 0.08}$. Remarkably, the
same power law variation of the density of dislocation with the
shear rate with an exponent $2/3$ is deduced from the SAXS data
and from the rheological results, thus providing convincing
support to our scenario for plastic deformation.

\begin{figure}
\includegraphics{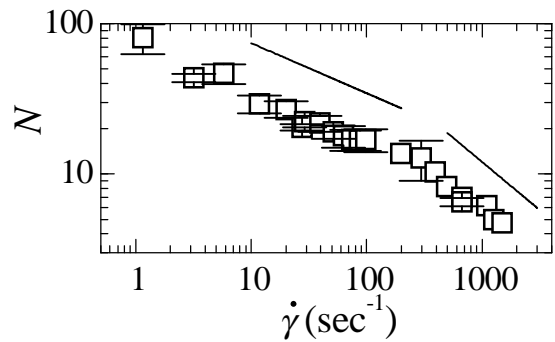}
\caption{Shear rate dependence of the normalized translational
correlation length, as measured from the broadening of the first
order Bragg peak. The two solid lines have slopes of $-1/3$ and
$-0.64$.} \label{FIG:4}
\end{figure}

The variation of $N=\xi/a_{0}$, the translational correlation
length (normalized by the lattice parameter $a_{0}$), with the
shear rate, below and above the melting transition, are reported
in Fig.\ \ref{FIG:4}. A continuous decrease of $N$ with
$\dot{\gamma}$, from $80$ to $4$, is measured. Our results clearly
show two regimes characterized on a log-log plot by a change of
slope at the shear rate $\dot{\gamma}_{c}$. The slope is equal to
$-1/3$ below $\dot{\gamma}_{c}$ and crosses over a more abrupt
variation, where a slope of $-0.64\pm0.06$ is measured. Although
the cross-over at the melting transition is smooth, it is
associated with a dramatic drop of viscosity and signs
unambiguously the melting transition. At $20 \, ^{\circ}\rm{C}$,
temperature at which the SAXS experiments are performed, both the
rheology and the SAXS experiments show that the shear melting
occurs for $\dot{\gamma}_{c}\simeq150 \rm \, {sec^{-1}}$. It is
associated with a critical normalized correlation length
$N_{c}=\frac{\xi_{c}}{a_{0}} \simeq 14 $ (Fig.\ \ref{FIG:4}),
which is interestingly of the same order of magnitude as the
correlation length measured at the melting of a $2$D colloidal
crystal \cite{2Dcolloids}. Additionally, for $b=a_{0}$, which are
presumably the most frequent dislocations, the numerical prefactor
$K$ in  Eq. (\ref{stressvsrho}) is equal to
$\frac{\sigma_{c}N_{c}}{G}$, where $\sigma_{c} \simeq 23 \,
\rm{Pa}$ (inset Fig.\ \ref{FIG:1}) is the critical stress at the
melting transition and $G \simeq 250 \rm \, {Pa}$ is the shear
modulus previously measured \cite{modulus}. One thus obtains $K
\simeq 1.3$, in excellent agreement with the value experimentally
found for metals \cite{HardeningExp}.

\begin{figure}
\includegraphics{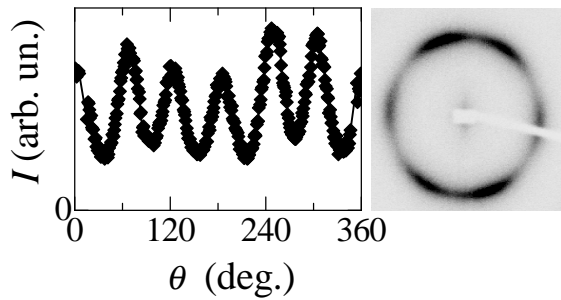}
\caption{(Right) 2D diffraction pattern of a sample submitted to a
shear rate  $\dot{\gamma}=390 \, \rm{sec}^{-1}$. The velocity is
perpendicular to the scattering plane. (Left) Azimuthal scan of
the $2\rm{D}$ diffraction pattern. The intensity $I$ is averaged
over an annulus of radius $q_{0}= 0.215 \, \rm{nm}^{-1}$ and width
$0.05 \, \rm{nm}^{-1}$ and is corrected to take into account the
curvature of the cell; once corrected, a six fold symmetry is
recovered with all maxima of roughly equal intensity.}
\label{FIG:5}
\end{figure}

Both the shear thinning regime and the SAXS scattering data can be
quantitatively and consistently interpreted as resulting from an
increase of the density of dislocations with the shear rate. We
therefore propose the shear melting be due to a proliferation of
dislocations. Our out-of-equilibrium results are in remarkable
analogy with the shear melting of FLL \cite{Corbino},
theoretically addressed when the shear is perpendicular to the
flux lines. In our case, however, the shear is oblique to the
tubes; the non trivial coupling between the gradual alignment of
the tubes and the increase of the density of dislocation with the
shear remains an open issue. Our experimental results compare also
naturally with those for the well documented melting of $2$D
solids \cite{2Dmelting}. However, because scattering experiments
do not provide direct imaging, no information about the spatial
distribution of the dislocations can be accessed, but only their
average density can be extracted from the data. As opposed to $2$D
solids, where direct imaging is possible
\cite{2Dcolloids,2Dobservations}, it is therefore delicate for
hexagonal columnar crystals to unambiguously characterize the
melting transition, and determine whether it is a one-stage first
order transition or a two-stage transition with an intermediate
line hexatic phase. Our data tend nevertheless to support the
latter scenario. Indeed, fig.\ \ref{FIG:5} shows the $2$D pattern,
obtained in tangential configuration for $\dot{\gamma} = 390 \,
\rm{sec}^{-1}$, that is above the melting transition. The
pronounced six-fold angular modulation observed on this pattern is
a signature of a line hexatic phase. Line hexatic phases have been
predicted \cite{Nelson,hexatictheory} and only very recently
observed by scattering techniques \cite{linehexatic,FLL}.
Interestingly, our experimental results also suggest the existence
under shear of an intermediate line hexatic phase, between the
columnar crystal and the liquid phase.

To conclude, we have shown that a mechanism of proliferation of
dislocations can qualitatively account for the melting of a
hexagonal columnar crystal under shear, which presumably occurs
through a two-step process with an intermediate line hexatic
phase. Our experimental results compare well with the
shear-melting of FLL and provide an original illustration of the
deep analogy between magnetic flux line lattices and columnar
liquid crystals. Recent developments in the simulations of
dislocation motions should render possible a numerical test of our
main findings and should provide decisive information about the
spatial organization of the dislocations in a steady-state regime.

The ESRF is acknowledged for financial support and provision of
synchrotron beam time and V. Urban for technical assistance during
the SAXS experiments. We thank M. Kl\'{e}man, M. Carmen-Miguel, D.
Lu and P. Olmsted for discussions, G. Porte for a critical reading
of the manuscript and one of the referee for pointing out Ref.
\cite{Corbino} to us.


\begin{references}
\bibitem{notepolycrystal} this is in principle true for a
monodomain; the extension to polydomain has to be taken with
caution.
\end{references}


\end{document}